\documentclass{article}
\usepackage{graphicx} 
\usepackage[hyphens,spaces,obeyspaces]{url} 
\usepackage{authblk} 
\usepackage{orcidlink} 
\usepackage{booktabs}
\usepackage{tabularx} 
\usepackage{float}
\usepackage{csquotes} 
\usepackage[natbibapa]{apacite} 

\title{Analysis of the Publication and Document Types in OpenAlex, Web of Science, Scopus, PubMed and Semantic Scholar}
\author[1]{Nick Haupka \orcidlink{0009-0002-6478-6789}}
\author[2]{Jack H. Culbert \orcidlink{0009-0000-1581-4021}}
\author[3]{Alexander Schniedermann \orcidlink{0000-0003-2132-7419}}
\author[1]{Najko Jahn \orcidlink{0000-0001-5105-1463}}
\author[2]{Philipp Mayr \orcidlink{0000-0002-6656-1658}}

\affil[1]{G\"ottingen State and University Library, University of G\"ottingen}
\affil[2]{GESIS -- Leibniz Institute for the Social Sciences}
\affil[3]{German Centre for Higher Education Research and Science Studies, DZHW}

\date{}

\begin{document}

\maketitle

\section*{Abstract}
The assignment of document and publication types in scholarly databases plays an important role in bibliometrics, for example in decision-making or university rankings. However, scholarly databases apply different curation strategies and taxonomies when classifying documents which makes it difficult to compare results from different database providers. In this study, the bibliometric databases OpenAlex, Web of Science, Scopus, PubMed and Semantic Scholar are used to analyse the extent of data variation and compare different approaches to taxonomy and data curation. Using a shared corpus of 9,575,603 publications from 2012 to 2022, we found large differences in the classification of document types such as research articles and editorials in these databases. We can also show that many of the records that lack a publication type in OpenAlex are classified as conference proceedings in Scopus and Semantic Scholar. 

\bigskip
\noindent\textbf{Keywords}: Publication Analysis, Publication Types, Coverage, Bibliometric databases, Open Scholarly Metadata

\section{Introduction}
The shift towards open science implies an emerging role of open metadata for scientometric analyses and has led to the development of open bibliometric data sources such as Semantic Scholar \citep{kinney_semantic_2023}, OpenAlex \citep{priem_openalex_2022} and Crossref \citep{hendricks_crossref_2020}. Consequently, an increasing body of research has focused on comparing these open resources to established commercial data sources in terms of coverage and quality \citep{waltman_special_2020,culbert_reference_2025}. In this large-scale study, we examine the classification of publication and document types in the open data sources OpenAlex, Semantic Scholar, and PubMed in comparison to the proprietary databases Scopus and Web of Science. 

Despite the varied typologies employed by these data sources, the extent of their differences remains unclear. Understanding these variations is crucial because the classification of publication and document types by bibliometric data sources influences the calculation of indicators.  For example, some bibliometric studies exclude editorials when calculating average citation rates for journals to ensure more meaningful and robust indicators. Conversely, the calculation of the original Journal Impact Factor can be manipulated by journals altering their document types, because this metric includes citations to all documents in the nominator but only articles and reviews in the denominator \citep{moed_improving_1995}. Other examples are university rankings such as the CWTS Leiden Ranking Open Edition, which only include a subset of publication and document types in the calculation \citep{van_eck_2024_13879947}.

Bibliometric data sources use different curation strategies to assign publication and document types. For instance, WoS tags a document as a review if it has been classified as a review by a journal or if the word \textit{review} is included in the title as well as in the text and has at least one reference\footnote{\url{https://webofscience.help.clarivate.com/en-us/Content/document-types.html}}. In contrast, PubMed's classification system also involves human indexers who apply the standardised Medical Subject Headings
(MeSH) \citep{van_buskirk_review_1984}. OpenAlex reuses and aggregates types from Crossref\footnote{\url{https://docs.openalex.org/api-entities/works/work-object\#type_crossref}} and has recently started to extend the existing classification to certain types such as reviews and preprints\footnote{\url{https://groups.google.com/g/openalex-users/c/YujaIIjY02A}}. The accuracy of these assignments is uncertain, with recent studies pointing to possible errors \citep{alperin_analysis_2024, visser_large-scale_2021}. An analysis by \citet{donner_document_2017} on document types in WoS and Scopus has shown that around 17 percent of publications in WoS had an incorrect document type classification. Similarly, \citet{mokhnacheva_document_2023} compared document types of 3,843 publications by Russian authors between 2010 and 2020 in Scopus and WoS and found differences in the assignment of document types from the databases and publisher websites. Using a semi-automated approach that includes a concordance matrix to identify misclassified document types, \cite{maisano_large-scale_2025} found about 2.3\% to 2.7\% of records in WoS and Scopus with inaccurate document type metadata.

In the following study, we first provide a conceptual overview of our interpretation of document and publication types. Building on these definitions, this study will then focus on a comparison of document and publication types between OpenAlex, Scopus, WoS, Semantic Scholar and PubMed. In addition to a general analysis of publication types, we examine and compare document types of specific items from journals from 2012 to 2022 to assess the quality of the classification of the corresponding data sources.

\section{Background}
In order to identify suitable research literature, research publications are classified according to their publication and document type. The publication type describes the location in which a text is published (e.g. a journal or a preprint server) whereas the document type represents the genre of a text (e.g. research article or review). The classification is based on specific characteristics, for example the length and scope of a text, the type of location in which it is published or the number of references \citep{harzing_document_2013} To systematise the problem of classifying research publications, we illustrate these two dimensions in more detail.
\begin{itemize}
    \item \textit{Publication or source types} represent the different venues in which document types are published. Such venues have been source types in traditional databases and can include periodic journals, monographs, edited volumes, conference proceedings and preprint servers. These venues have been the subject to the curating procedures of databases such as WoS, leading to the inclusion of items from selected sources covered by the policies of each database. Indexing practices can vary, for example, Scopus only indexes serial publications assigned to an International Standard Serial Number (ISSN), as well as one-off conferences and one-off books \footnote{\url{https://web.archive.org/web/20240527182403/https://assets.ctfassets.net/o78em1y1w4i4/EX1iy8VxBeQKf8aN2XzOp/c36f79db25484cb38a5972ad9a5472ec/Scopus_ContentCoverage_Guide_WEB.pdf}}. Additionally, these sources must comply to the content coverage policy in Scopus. In contrast, OpenAlex does not select journals \footnote{\url{https://docs.openalex.org/api-entities/sources}}, which results in a high number of journals that are only present in the OpenAlex database, but not in WoS or Scopus \citep{maddi_geographical_2025}. 
    \item \textit{Document types} represent the types or genres of individual texts and are usually included as metadata in databases. Some document types, such as editorials or letters, are similar in their textual characteristics and scholarly functions, because they are rather short texts written in a more commentary-style. A larger subset comprises review articles that do not present original research findings, but instead summarise and compile previous results to form a scientific consensus on a specific field or topic. However, the majority of document types are common journal articles containing the results and findings of most research activities, or provide theoretical contributions and conceptual essays. Articles published in journals are also considered by some to be the gold standard for bibliometric research and research evaluation \citep{moed_measuring_2004}
\end{itemize}

These two dimensions represent different textual or social characteristics of documents in science and can provide complex classification problems that can lead to confusion and errors in bibliometric research. Document types can be confused with publication types or venues in bibliometric studies or used synonymously \citep{scheidsteger_comparison_2023}. This is further complicated as document type classifications can be derived from the publication venue. Preprints may be classified solely by their repository as a source venue. Likewise, journal articles published in review journals may be classified as reviews. Further, classifications can be derived from disciplinary definitions that make them less comparable in cross-disciplinary analyses \citep{sigogneau_2020}.

As indicated above, scholarly databases proceed differently when classifying document and publication types. The providers Elsevier (Scopus) and Clarivate (WoS) employ a curated list of venues that are indexed in the respective databases. The classification of document types contained therein is carried out by an editorial team \footnote{\url{https://clarivate.com/academia-government/scientific-and-academic-research/research-discovery-and-referencing/web-of-science/web-of-science-core-collection/editorial-selection-process/}}\footnote{\url{https://www.elsevier.com/products/scopus/content}}. The database PubMed consists of sources indexed in MEDLINE and journals deposited in PubMed Central \footnote{\url{https://www.nlm.nih.gov/bsd/difference.html}}. PubMed only features publications from biomedical research. With MeSH, PubMed features a multi-purpose document type classification system that was originally based on human indexing and only recently shifted to automated indexing \citep{mork_nlm_2013}. The corpus of Semantic Scholar comprises sources from many different vendors. Among these are non-profit organisations such as Crossref, preprint servers like arXiv, academic publishers and sources found by a web crawler \citep{kinney_semantic_2023}. Similarly, OpenAlex collects items from many different sources such as preprint servers, repositories, publishers and websites \citep{priem_openalex_2022}. The information about these sources comes from Crossref, the ISSN Network and the Microsoft Academic Graph (MAG) \footnote{\url{https://docs.openalex.org/api-entities/sources}}. The classification of document types is also based on information from Crossref \footnote{This holds true for data from OpenAlex that was published before 2024. Since 2024, OpenAlex applies a custom document type classification (See also \cite{haupka_recent_2024})}.

The different approaches of the operators can lead to distortions in bibliometric analyses. For example, OpenAlex and WoS differ in the number of research articles, because WoS classifies several articles in OpenAlex as editorials, letters and book reviews \citep{mongeon_investigating_2025}. Further, document types are sometimes not supported by a database, e.g. the document type data paper exists in Scopus and WoS but not in Dimensions and OpenAlex \citep{jiao_how_2023}. Sometimes items in databases can not be found, because a database lacks metadata information about the document type. This is the case, for example, with Semantic Scholar where only 41\% of records are assigned to a document type \citep{delgado-quiros_completeness_2024}. This problem can be extended to public repositories, which in some cases aggregate metadata from several thousand providers, whereby the quality of the metadata is not always assured \citep{charalampous_classifying_2017}. Occasionally, the same or identical text can occur in multiple different venues. Although this form of multiple publication can be legitimate, it leads to multiple and conflicting document type classifications. For example, the addition or removal of references during the peer review process may turn a preprinted review into a published article if the classification system is solely based on the number of references. Similarly, systematic reviews in medicine can be (re-)published as updated or abbreviated versions \citep{bashir_sr_update_2018}, and method papers or guidelines are translated into other languages.

\section{Data and Methods}
\subsection{Data}\label{sec:data}
In this work, publication and document types from Openalex (August 2023 snapshot), Web of Science (July 2023 snapshot), Scopus (July 2023 snapshot), PubMed (December 2022 snapshot) and Semantic Scholar (September 2023 snapshot) were analysed. The WoS data includes the collections: Science Citation Index Expanded (SCI), Social Sciences Citation Index (SSCI), Arts and Humanities Citation Index (AHCI), Conference Proceedings Citation Index-Science (ISTP) and Conference Proceedings Citation Index-Social Sciences \& Humanities (ISSHP)\footnote{The Preprint Citation Index was not used in this study due to license restrictions.}. 

We restricted the data to the publication years 2012 to 2022 and to publications with a Digital Object Identifier (DOI). The publication year from OpenAlex was used, unless otherwise stated. The publication year of a publication may differ between the investigated databases, depending on which version of a work is indexed in the database \citep{whisperer_when_2022, delgado-quiros_completeness_2024}. Because the selected PubMed snapshot is from December 2022, not all publications in PubMed from the publication year 2022 were fully covered. All data were analysed within a custom SQL database environment maintained by the German Competence Network of Bibliometrics \citep{schmidt_data_2024}. 

For comparison, we used the DOI instead of the PubMed identifier (PMID) to match records from the respective databases. Since the inception of versioning for PMIDs, the ID no longer points to a unique item, i.e., subsequent stages of the same preprint can obtain the same PMID but differ in the preprint version \citep{torre_versioning_2012}. As version information is currently not indexed by WoS or Scopus, matching by PMID would cause inflated numbers of matches. We normalised DOIs for items in each database beforehand. This step included the removal of URL prefixes and adjusting character strings to lowercase.

The number of publications for each database analysed can be found in Table \ref{tab:datasources}. The table also shows the number of joint publications between the databases OpenAlex, Scopus, WoS, PubMed and Semantic Scholar. Following \citet{culbert_reference_2025}, this intersection is referred to as a shared corpus.

\begin{table}[h]
\centering
\begin{tabularx}{\textwidth}{lX}
\toprule
\textbf{Data source} & \textbf{Number of publications} \\ 
\hline
OpenAlex & 69,456,021 \\
Scopus & 31,922,514 \textbf{(96\% included in OpenAlex)} \\
Web of Science & 23,540,852 \textbf{(99\% included in OpenAlex)} \\
Semantic Scholar & 56,279,413 \textbf{(97\% included in OpenAlex)} \\
PubMed & 12,681,219 \textbf{(99\% included in OpenAlex)}  \\
\hline
Shared Corpus & 9,575,603 \textbf{(Only journal items)} \\
\bottomrule
\end{tabularx}
\caption{Data Sources used in this study. The number of publications is limited to those records with publishing year between 2012 and 2022 inclusively. No restriction was placed on the publication or document type (with the exception of the shared corpus, which includes only journal items).}
\label{tab:datasources}
\end{table}

Focussing on the metadata of publication and document types in the the databases analysed, it shows that this process is handled differently in each data source. In PubMed, publication types are not clearly separable from document types because the database only provides a single metadata field that can cover publication types, document types, study types and even funding information. In OpenAlex and Scopus, one publication and one document type were assigned to each item. In the case of OpenAlex, the venue type specification in the primary location was considered. The primary location is where the best copy of a work can be found (closest to the version of record)\footnote{\url{https://docs.openalex.org/api-entities/works/work-object\#primary_location}}. In Semantic Scholar and WoS, items can be assigned to multiple document types. These were all taken into account in this study and were not reduced to one type, which means that the calculated numbers in the Tables \ref{tab:types_in_journals} and \ref{tab:reclassification} can differ between the data sources when the shared corpus is analysed. WoS and Semantic Scholar specified one publication type for each publication. 

\subsection{Methodology}\label{sec:methodology}
In the first part of the following analysis, we examined publication types in OpenAlex in comparison to Scopus, WoS and Semantic Scholar (see Section \ref{sec:pub_types}). In the second part, we investigated document types of journal publications in OpenAlex, Scopus, WoS, Semantic Scholar and PubMed (see Section \ref{sec:doc_types}). To determine the proportion of editorial and research publications in the respective databases (see Table \ref{tab:reclassification}), we categorised document types into three groups:

\textbf{Research Discourse}: This includes articles, reviews and clinical studies.\footnote{We considered retracted publications as research discourse in this study.
}

\textbf{Editorial Discourse}: This covers letters, editorials, comments and retraction notes.

\textbf{Not Assigned}: All works that cannot be assigned to either research or editorial discourse were categorised here.

There are some specific aspects of the OpenAlex snapshot used, which are briefly described here:

In July 2023, OurResearch implemented a new item classification, replacing the previously used item classification of Crossref in OpenAlex\footnote{\url{https://docs.openalex.org/api-entities/works/work-object\#type}}. In this process, the types \textit{journal-article}, \textit{proceedings-article}, and \textit{posted-content} were combined into the type \textit{article}, which makes it difficult to differentiate between conference articles, preprints, and journal articles. As this transition has not yet been fully completed in the used snapshot, we have manually reclassified the remaining old types using the OpenAlex procedure\footnote{\url{https://github.com/ourresearch/openalex-guts/}}.

Within the OpenAlex database, retractions were not assigned distinct document types. Instead, OpenAlex provided a designated field to label retractions (\textit{is\_retracted})\citep{hauschke_non-retracted_2025}. Because retractions are not separated into a unique category within OpenAlex, they are likely included in most analyses by default. For this analysis, we ignored the information from OpenAlex in the fields \textit{is\_paratext} and \textit{is\_retracted}. 

We also reclassified types derived from PubMed in Table \ref{tab:types_in_journals}. Because PubMed provides multiple document types to one single item, we reduced the multi-assignment by employing a hierarchy that ranks the individual document types. For example, we ranked research-related types highest, followed by editorials, letters, news items, review articles, and journal articles (see Table \ref{tab:pm_hierarchy} for the complete scheme). After ranking, only the highest-ranked type was retained for each item.

\section{Results and Discussion}

\subsection{Overview}
First, we identified the different typologies in our data sources and assess how comprehensively each category is covered (see Table \ref{tab:datatypo}). We then calculated the coverage ratio separately for publication and document types. 

We found that, compared to the proprietary providers, the open databases have a lower coverage of publication and document types. Scopus and WoS achieve full coverage of document types. In WoS, each item is also assigned a publication type. This value is similarly high in Scopus (99.97\%). However, both databases differ in terms of typology. WoS offers a much more granular document type classification system compared to Scopus, with 87 categories versus  18. 

Focusing on the open databases, it can be seen that OpenAlex and PubMed also nearly have 100\% of the items assigned with a document type. However, the proportion of items with a publication type was lower than that of Scopus and WoS (86\% of items in OpenAlex have a publication type). No publication types were separately specified in PubMed. The number of document types in PubMed is similar to Web of Science (79 document types). 

Overall, Semantic Scholar had the lowest coverage, providing two publication and 12 document types. Less than half of the items indexed in Semantic Scholar were assigned to a publication type, and 37\% to a document type. 

\begin{table}[h]
\centering
\makebox[\textwidth]{
\begin{tabularx}{1.3\textwidth}{lXXXX}
\toprule
\textbf{Data Source} & \textbf{Publication Type \newline Typologies} & \textbf{Publication Type \newline Coverage} & \textbf{Document Type \newline Typologies} & \textbf{Document Type \newline Coverage}\\ 
\hline
OpenAlex & 5 & 85.85\% & 16 & 100\% \\
Scopus & 7 & 99.97\% & 18 & 100\%  \\
Web of Science & 3 & 100\% & 87 & 100\% \\
Semantic Scholar & 2 & 43.60\% & 12 & 37.04\% \\
Pubmed & / & / & 79 & 99.99\%  \\
\bottomrule
\end{tabularx}}
\caption{Size of publication type and document type typologies per data source. The coverage ratio is calculated separately for publication and document types.}
\label{tab:datatypo}
\end{table}

\subsection{Publication types}\label{sec:pub_types}
\begin{figure}[H]
\centering
\makebox[\textwidth]
    {\includegraphics[width=14cm]{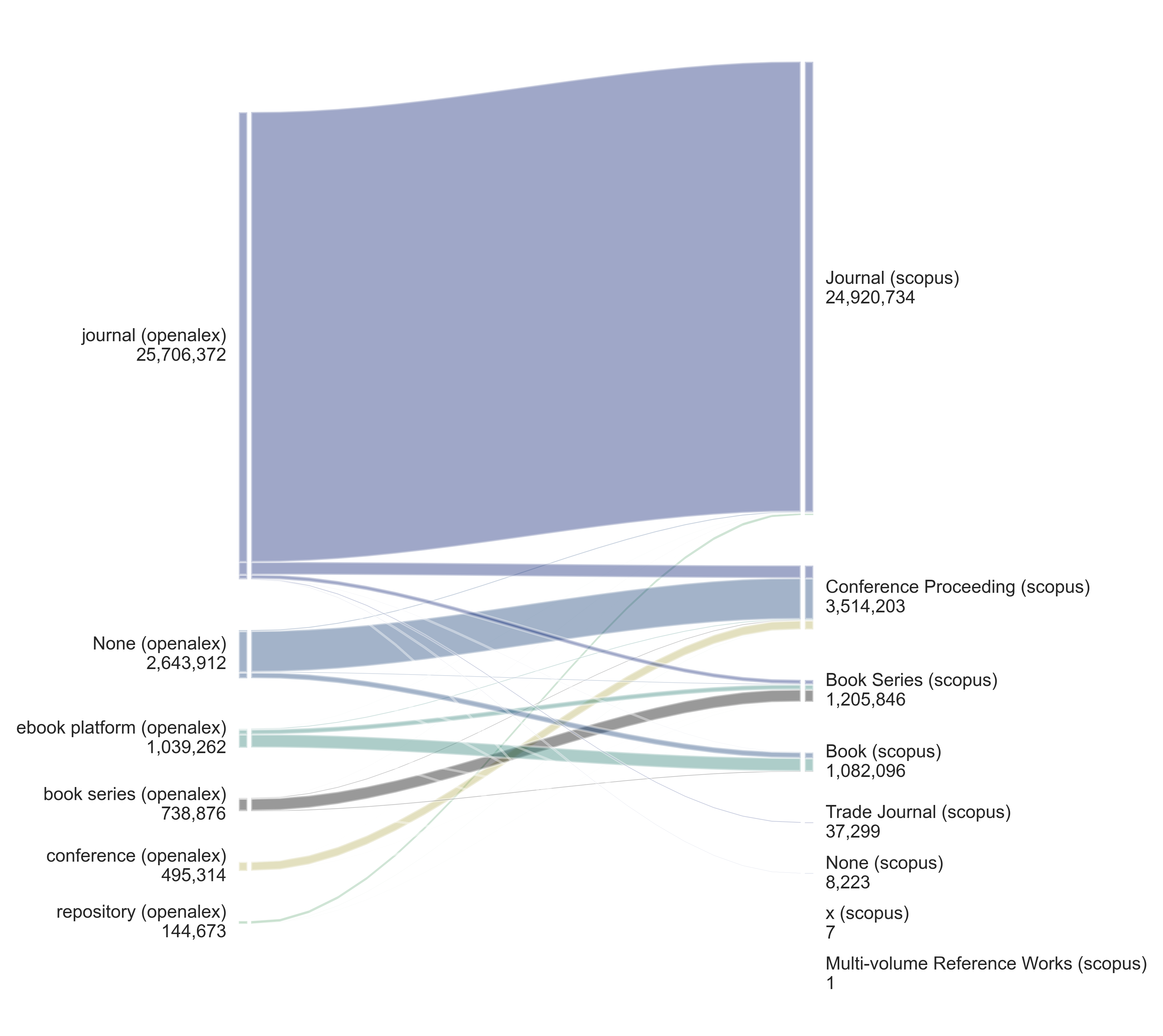}}
\caption{\label{fig:oal_scp_venues}Publication type classification between OpenAlex and Scopus. The intersection between Scopus and OpenAlex is limited to those records with publishing year between 2012 and 2022 inclusively. The source type \enquote{x} is an artefact from the Scopus XML raw files.}
\end{figure}

\begin{figure}[H]
\centering
\makebox[\textwidth]
    {\includegraphics[width=14cm]{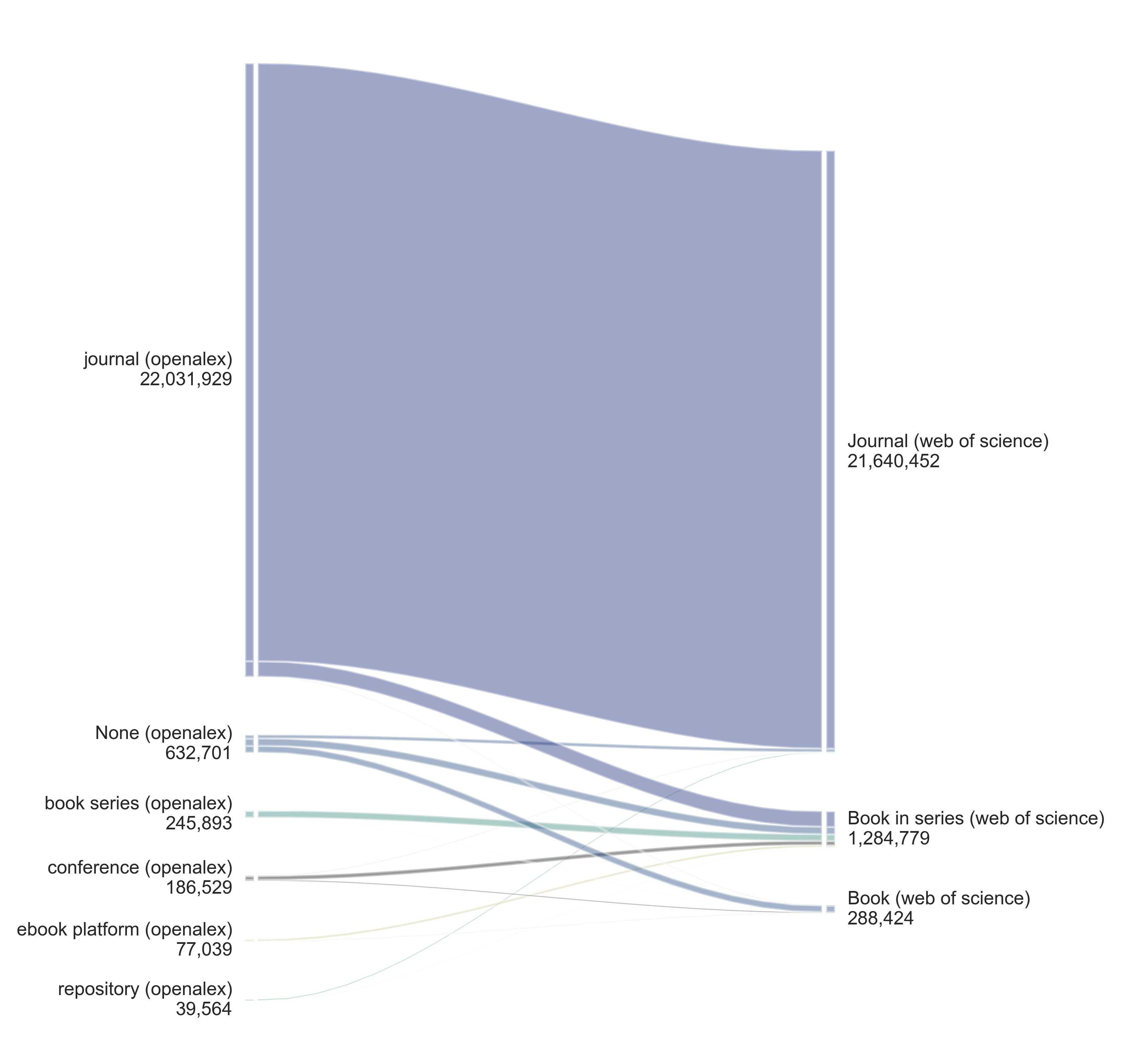}}
\caption{\label{fig:oal_wos_venues}Publication type classification between OpenAlex and WoS. The intersection between WoS and OpenAlex is limited to those records with publishing year between 2012 and 2022 inclusively.}
\end{figure}

\begin{figure}[H]
\centering
\makebox[\textwidth]
    {\includegraphics[width=14cm]{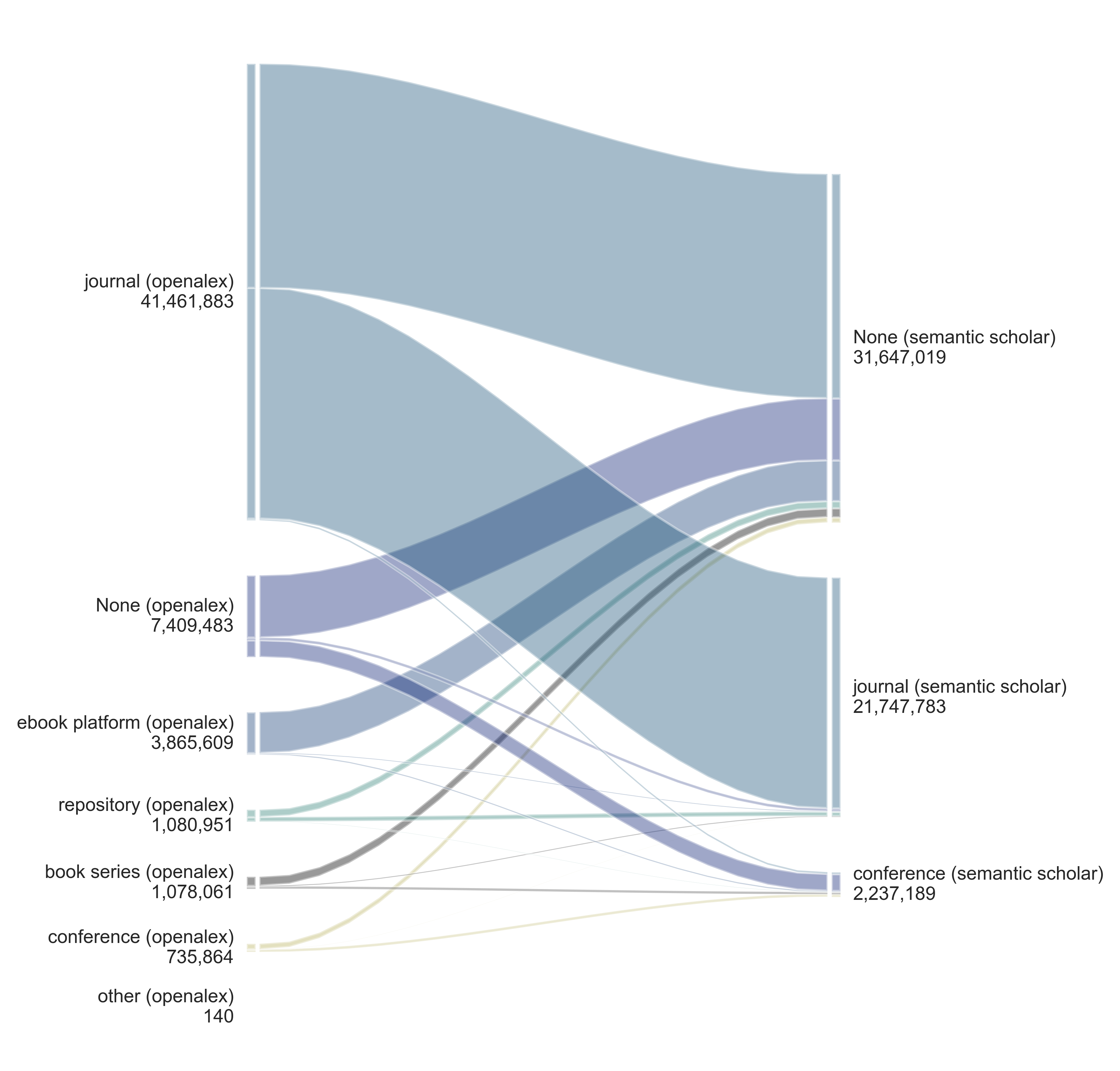}}
\caption{\label{fig:oal_s2_venues}Publication type classification between OpenAlex and Semantic Scholar. The intersection between Semantic Scholar and OpenAlex is limited to those records with publishing year between 2012 and 2022 inclusively.}
\end{figure}

The analysis of publication types illustrated that the various database providers adopt different approaches to classifying venues (see Figure \ref{fig:oal_scp_venues}, \ref{fig:oal_wos_venues}, \ref{fig:oal_s2_venues}). In addition,  providers have different focuses on certain types of publication. For Figures \ref{fig:oal_scp_venues}, \ref{fig:oal_wos_venues} and \ref{fig:oal_s2_venues}, we have considered the intersection of a selected database with OpenAlex for the publication years 2012 to 2022. This means that publications and the number of publication types that are not included in one of the databases are not shown. The comparison is a 1:1 allocation, because only one publication type was present to each of the databases examined. 

Journals comprised the largest share in all databases. This was followed by conference proceedings and books. A large proportion of items was not assigned to a venue, in particular in Semantic Scholar. 

Publications that were not assigned to a venue in OpenAlex were largely classified as conference proceedings in the other investigated databases. Conference proceedings have been included in WoS, however, the publication type field lacks an explicit designation for this type. Conference proceedings contained in WoS are indexed as publication type \textit{book}, \textit{book in series} or \textit{journal}.

In Semantic Scholar books are not included as a publication type so that books from OpenAlex cannot be matched to any venue. Items that are assigned to the venue type repository in OpenAlex are generally less covered in Web of Science and Scopus where they are categorised as a journal. Compared to Semantic Scholar, more repositories from OpenAlex are covered and match either to the type journal (410,444 items) or no venue at all (653,592 items). 

\subsection{Document Types}\label{sec:doc_types}
\begin{figure}[H]
\centering
\makebox[\textwidth]
    {\includegraphics[width=14cm]{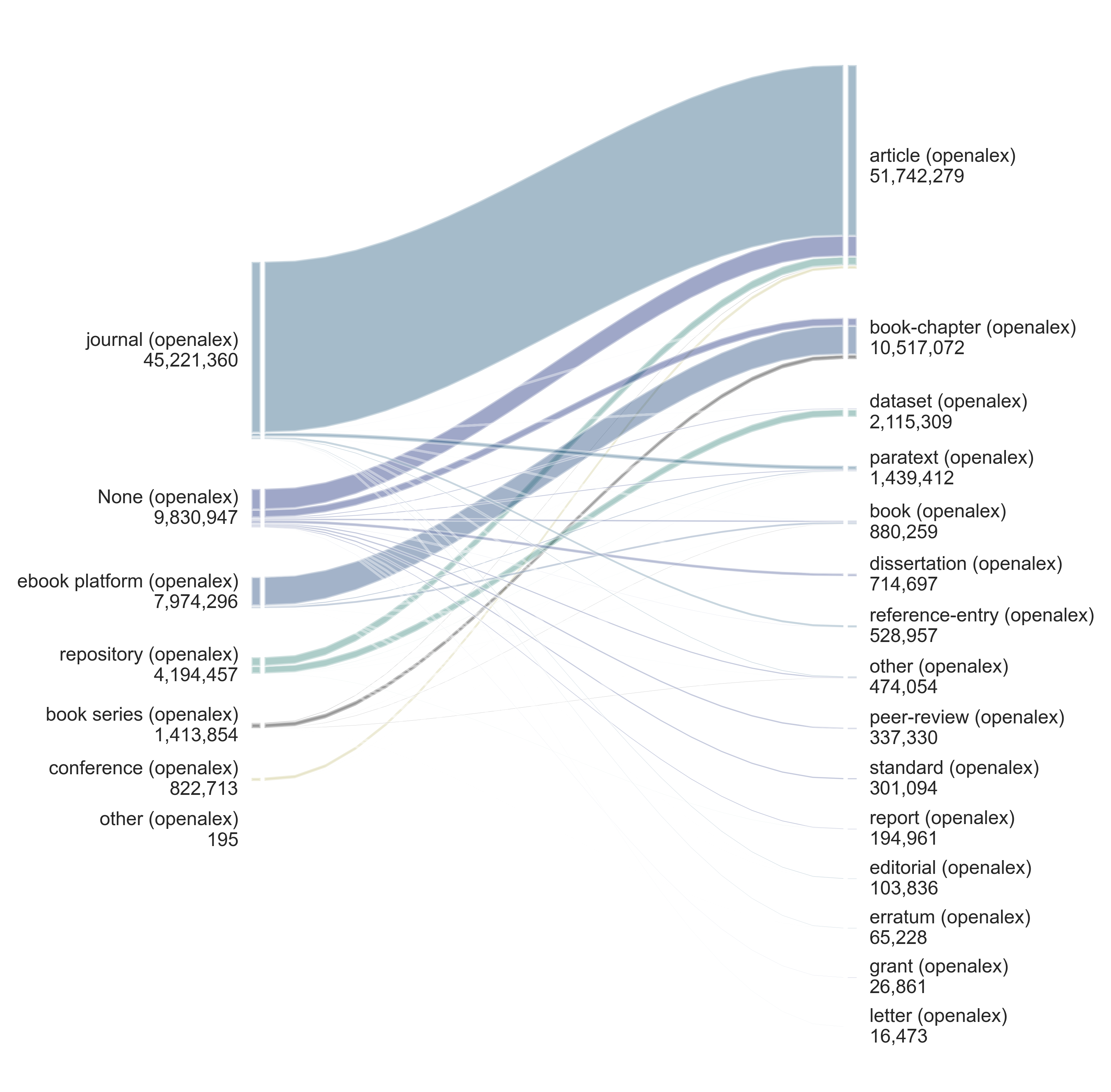}}
\caption{\label{fig:oal_type_venue}Publication types and their corresponding document types in OpenAlex. Data is limited to those records with publishing year between 2012 and 2022 inclusively.}
\end{figure}

In this section, we first demonstrate how the different item types correspond to the venue types in OpenAlex (see Figure \ref{fig:oal_type_venue}). This is followed by a comparison of document types from journal publications in the databases OpenAlex, Scopus, WoS, PubMed and Semantic Scholar (see Table \ref{tab:types_in_journals}).

Starting with Figure \ref{fig:oal_type_venue}, we can see that the majority of items classified as articles in OpenAlex had the publication type journal. Approximately five million items classified as article were not assigned to a publication type. However, Figures \ref{fig:oal_scp_venues} and \ref{fig:oal_s2_venues} suggest that these could be conference articles, as a certain proportion of the items that are not assigned to a publication type in OpenAlex have been classified as conference publications by Scopus and Semantic Scholar. Around two million articles have the venue type repository. According to OurResearch, these are preprints\footnote{\url{https://github.com/ourresearch/openalex-guts/}}, because MAG, as a precursor to OpenAlex, also covered preprint servers as venues \citep{xie_is_2021} and their items were labeled as journal articles in OpenAlex \citep{scheidsteger_comparison_2023}. In fact, when counting items from repositories per source title since 2020, most stem from arXiv, SSRN, or bioRxiv. Figures \ref{fig:oal_scp_venues} and \ref{fig:oal_s2_venues} demonstrate that these have the publication type journal in Scopus and Semantic Scholar. The publication type book and book series was represented by around 9 million items. The number of publication type classified as \textit{conference} was determined to be almost 1 million for the period 2012 to 2022.

Table \ref{tab:types_in_journals} presents the document types of items published in journals between 2012 and 2022 in the above mentioned shared corpus (see Section \ref{sec:data}). It can be seen that journal articles constituted the largest share in all data sources. OpenAlex classified over 99\% of the items as journal articles. This proportion was lower in Scopus (80\%), WoS (78\%), Semantic Scholar (73\%) and PubMed (75\%). Except for OpenAlex, all databases have classified reviews. With around 9\%, reviews  constituted the second largest share. Only Semantic Scholar had a larger share of reviews (15\%). Editorial materials such as editorials, errata or letters are classified in all databases. Here, OpenAlex, classified around 0.3\% of items as editorial material. In Scopus, around 4\% of items were assigned to the document type letter, 2\% note and 2\% editorials. In WoS, the proportion of letters was similar (4\%). The proportion of editorials was slightly higher in WoS than in Scopus (around 7\%). The proportion of letters in PubMed and Semantic Scholar were similar to those in WoS and Scopus (4\%).

\begin{table}[h]
\centering
\makebox[\textwidth]{
\begin{tabularx}{1.6\textwidth}{XX|XX|XX|XX|XX}
\toprule
\multicolumn{2}{c|}{\textbf{OpenAlex}} & \multicolumn{2}{c|}{\textbf{Scopus}} & \multicolumn{2}{c|}{\textbf{WoS}} & \multicolumn{2}{c|}{\textbf{Semantic Scholar}} & \multicolumn{2}{c}{\textbf{PubMed}} \\
\hline
Type & n & Type & n & Type & n & Type & n & Type & n \\
\hline
article & 9,528,567 \textbf{(99.51\%)} & Article & 7,618,472 \textbf{(79.51\%)}& Article & 7,535,712 \textbf{(77.77\%)}& Journal Article & 8,836,640 \textbf{(72.95\%)}& Journal Article & 7,178,813 \textbf{(74.96\%)}\\
editorial & 20,316 \textbf{(0.21\%)} & Review & 923,486 \textbf{(9.64\%)} & Review & 878,361 \textbf{(9.07\%)} & Review & 1,765,638 \textbf{(14.58\%)} & Review & 878,361 \textbf{(9.19\%)}\\
erratum & 15,201 \textbf{(0.15\%)} & Letter & 358,420 \textbf{(3.74\%)} & Editorial Material & 637,569 \textbf{(6.58\%)} & Letters And Comments & 455,431 \textbf{(3.76\%)} & Letter & 352,031 \textbf{(3.68\%)}\\
letter & 9,712 \textbf{(0.1\%)} & Note & 251,305 \textbf{(2.62\%)} & Letter & 375,866 \textbf{(3.88\%)} & Study & 410,735 \textbf{(3.40\%)} & Clinical Trial & 293,069 \textbf{(3.06\%)}\\
paratext & 636 \textbf{(0.01\%)} & Editorial & 217,034 \textbf{(2.27\%)} & Correction & 93,702 \textbf{(0.97\%)} & Case \newline Report & 263,532 \textbf{(2.18\%)} & Case \newline Report & 279,377 \textbf{(2.92\%)}\\
reference-entry & 520 \textbf{(0.01\%)} & Erratum & 99,979 \textbf{(1.04\%)} & Proceedings Paper & 70,846 \textbf{(0.73\%)} & Editorial & 215,889 \textbf{(1.78\%)} & Editorial & 229,707 \textbf{(2.4\%)}\\
report & 1 \textbf{(0\%)} & Conference Paper & 53,858 \textbf{(0.56\%)} & News Item & 37,479 \textbf{(0.39\%)} & Clinical Trial & 71,035 \textbf{(0.59\%)} & Meta Analysis & 198,037 \textbf{(2.07\%)} \\
Other & 654 \textbf{(0.01\%)} & Other & 59,403 \textbf{(0.62\%)} & Other & 60,029 \textbf{(0.62\%)} & Other & 93,663	 \textbf{(0.77\%)} & Other & 165,334 \textbf{(1.73\%)} \\
\bottomrule
\end{tabularx}}
\caption{Document type classification between OpenAlex, Scopus, Web of Science, Semantic Scholar and PubMed. The table contains journal publications between 2012 and 2022. Only publications that are present in all five databases are considered.}
\label{tab:types_in_journals}
\end{table}

In order to compare the ratio of research and non-research publications in journals between the different data sources, document types of the respective databases were reclassified (see Table \ref{tab:reclassification}). Here, we use the classes \textit{research discourse} and \textit{editorial discourse} (see also  Section \ref{sec:methodology}). Types that cannot be assigned to either class (such as books or dataset) are not assigned to any class (not assigned). Only items that were published in journals between 2012 and 2022 and can be found in all five databases were considered in the following table. 

From Table \ref{tab:reclassification}, it can be seen that OpenAlex classified only a low proportion of items as editorial discourse (0.5\%). The proportion of editorial material is higher in Scopus (10\%), WoS (12\%), Semantic Scholar (6\%) and PubMed (8\%). The proportion of research items, in turn, is particularly high in OpenAlex with over 99\%, which indicates that OpenAlex classifies a large number of non-research items as research publications. 

\begin{table}[h]
\centering
\makebox[\textwidth]{
\begin{tabularx}{1.4\textwidth}{l|X|X|X|X|X}
\toprule
{} & {\textbf{OpenAlex}} & {\textbf{Scopus}} & {\textbf{WoS}} & {\textbf{Semantic Scholar}} & {\textbf{Pubmed}} \\
\hline
Research Discourse & 9,528,567 \textbf{(99.51\%)} & 8,542,568 \textbf{(89.15\%)} & 8,417,586 \textbf{(86.87\%)} & 11,408,341 \textbf{(94.19\%)} & 11,362,371 \textbf{(88.58\%)} \\
Editorial Discourse & 45,865 \textbf{(0.48\%)} & 926,738 \textbf{(9.67\%)} & 1,150,387 \textbf{(11.87\%)} & 703,573 \textbf{(5.81\%)} & 985,505 \textbf{(7.68\%)} \\
Not assigned & 1,175 \newline \textbf{(0.01\%)} & 112,651 \textbf{(1.18\%)} & 121,591 \textbf{(1.25\%)} & 649 \newline \textbf{(0.01\%)} & 479,936 \textbf{(3.74\%)} \\
\bottomrule
\end{tabularx}}
\caption{Classification of document types into research discourse and editorial discourse in OpenAlex, Scopus, Web of Science, Semantic Scholar and PubMed. The table contains journal publications between 2012 and 2022. Only publications that are present in all five databases are considered.}
\label{tab:reclassification}
\end{table}

\subsection{Discussion}
This study compared the publication and document types in OpenAlex with those in Scopus, WoS, Semantic Scholar, and PubMed. The results show structural differences in the curation strategies and taxonomies of the databases analysed. Firstly, we identified a gap in the coverage of publication type metadata (see Figure \ref{tab:datatypo}). While almost every publication is assigned to a publication type in WoS and Scopus, only 44\% of publications are assigned to a publication type in Semantic Scholar and around 86\% in OpenAlex. One possible explanation for the low coverage of publication types in Semantic Scholar is that Semantic Scholar extracts metadata mainly from the web, where information is not fully available \citep{delgado-quiros_completeness_2024}. We also encountered problems with the classification of conference proceedings in OpenAlex. A large proportion of publications with a missing publication type in OpenAlex are classified as conference proceedings in Scopus (see Figure \ref{fig:oal_scp_venues}). This observation is also made when comparing publication types in OpenAlex and Semantic Scholar (see Figure \ref{fig:oal_s2_venues}). A closer look at the classification has also shown that OpenAlex sometimes classifies venues as journals when they should actually be categorised as conference proceedings. For example, the venue \enquote{The Proceedings of the Annual Convention of the Japanese Psychological Association} is classified as a journal in OpenAlex, but as a conference proceeding on the publishers website \footnote{\url{https://openalex.org/sources/s4210219382}}. Another example is the venue \enquote{EPJ Web of Conferences} \footnote{\url{https://openalex.org/sources/s19068271}}. Besides that we found a large overlap in the classification of books between OpenAlex, Scopus and WoS. This could possibly be due to the ISBN, which makes it easier to classify resources as books, whereas ISSN can be used for journals but also for other periodical resources.

The analysis of the document types revealed a high degree of completeness in the coverage of metadata in OpenAlex, Scopus, WoS and PubMed (see Figure \ref{tab:datatypo}). Only publications in Semantic Scholar have a low coverage of document type metadata (about 37\%). Again, a possible explanation is that OpenAlex, Pubmed, Scopus and WoS rely more on publisher information and additional curation efforts, while Semantic Scholar mainly aggregates data from the web \citep{delgado-quiros_completeness_2024}. In the case of OpenAlex, Crossref is the main provider of bibliometric metadata. Crossref acts as a collection point for research publications worldwide, which are usually enriched with metadata by publishers and institutions \citep{hendricks_crossref_2020}. In contrast to OpenAlex (and probably Semantic Scholar), PubMed not only uses publisher information, but also involves human reviews and curation when indexing publications \footnote{\url{https://www.nlm.nih.gov/bsd/indexfaq.html}}\footnote{\url{https://pubmed.ncbi.nlm.nih.gov/help/}}. Because PubMed performs a manual validation of the document types, it achieves a high degree of accuracy in the assignment of document types to publications \citep{yeung_comparison_2019}. This could explain, why PubMed only classifies about 89\% of publications as research publications, Semantic Scholar 94\% and OpenAlex 99\% in our analysis (see Table \ref{tab:reclassification}). 
Apart from this, the typology in PubMed is much more comprehensive compared to the other databases and covers medical publication types such as clinical trials that are not included in the other databases, with the exception of Semantic Scholar (which includes the document types CaseReport and ClinicalTrial). In Scopus, case reports are considered as research articles \footnote{\url{https://www.elsevier.com/products/scopus/content}}. WoS also has a detailed document type typology with over 85 different labels. However, the typology does not contain any assignments of publications to specific medical document types. Instead, WoS is the only data source that explicitly covers the document type book review. In OpenAlex, book reviews are usually classified as articles and in Semantic Scholar they are predominantly classified as reviews. This finding is consistent with \cite{mongeon_investigating_2025}, who also observed that book reviews are largely assigned to the document type article or review in OpenAlex. In Scopus, book reviews are explicitly listed as document types that are not recorded in the database \footnote{\url{https://www.elsevier.com/products/scopus/content}}. Another observation from this study is that although editorials and letters are classified in all databases, the number of these publications is particularly low in OpenAlex (only about 0.3\%), while the value is similar in Scopus, Semantic Scholar and PubMed (in sum about 5 to 6\%). WoS classifies more records as editorials and letters (in sum about 10\%). The high proportion of articles in OpenAlex can be attributed to Crossref, as OpenAlex reuses publication metadata from this data source. Publication metadata from Crossref, in turn, is largely supplied by publishers \footnote{\url{https://crossref.gitlab.io/knowledge_base/docs/topics/content-types/}}. Semantic Scholar also reuses Crossref metadata, however, it is not clear to which extent \citep{kinney_semantic_2023, delgado-quiros_why_2024}.

\section{Conclusion}
This paper demonstrates inconsistencies in publication and document type classification across bibliometric databases. This problem stems from the lack of a standardised system for classifying publications. Each database utilises its own typologies and approaches which makes overarching comparisons difficult (see Table \ref{tab:reclassification}). Furthermore, the results indicate that the databases Web of Science, Scopus and PubMed classify more publications as editorial content compared to OpenAlex and Semantic Scholar. However, the comparison of publication types shows that, with the exception of Semantic Scholar, all databases use a similar classification of venues, with a large overlap between OpenAlex, Scopus and Web of Science. The results also show that OpenAlex and Semantic Scholar lack metadata on publication types, which can lead to a lower recall when searching for records from these databases. Future studies should be aware of these structural and methodological differences when interpreting results from bibliometric databases.

\section{Limitations}
Since the submission of this study, OpenAlex has made some improvements to the existing document type classification that could no longer be taken into account in this analysis. In May 2024, the document types preprint, libguides, supplementary-materials and reviews were added. The document type retraction was added in July 2024. Since July 2024, OpenAlex also uses PubMed to reclassify document types. An updated analysis of the document types in OpenAlex, Scopus and WoS can be found in \cite{haupka_recent_2024}. The results indicate an improvement in document type classification when comparing to Web of Science and Scopus. Note that the changes introduced by OpenAlex in the classification of document types make it difficult to reproduce our findings, as OurResearch only provides access to the latest OpenAlex snapshot.
Other limitations in this study include the selection of publication types. In this analysis, we only used the publication source, which was identified by the databases as the closest version to the original. However, it is common for publications to have multiple publication locations. A further limitation is the inclusion of PubMed data in the shared corpus, which results in an over-representation of publications from the life sciences, as PubMed restricts its data to certain types of topics. Finally, the lack of ground truth makes it difficult to examine the accuracy of publication and document types in bibliometric databases on a large scale.

\newpage
\section*{Funding Information}
This work was funded by the Federal Ministry of Education and Research (grant funding number: 16WIK2301B / 16WIK2301E, The OpenBib project; grant funding number: 01PU17017, The FuReWiRev project). We acknowledge the support from the Federal Ministry of Education and Research, Germany under grant number 01PQ17001, the Competence Network for Bibliometrics.

\section*{Code Availability}
The Jupyter notebook containing the source code analysis can be found on GitHub: \url{https://github.com/naustica/openalex\_doctypes}.

\section*{Author Contributions}
Nick Haupka: Conceptualization, Data Curation, Formal Analysis, Investigation, Methodology, Software, Validation, Visualization, Writing—Original Draft, Writing—Review and Editing. \\
Jack Culbert: Data Curation, Investigation, Validation, Writing—Review and Editing.\\
Alexander Schniedermann: Data Curation, Writing—Review and Editing. \\
Najko Jahn: Funding acquisition, Project administration, Supervision, Writing—Review and Editing.\\
Philipp Mayr: Funding acquisition, Project administration, Supervision, Writing—Review and Editing.\\
\newpage
\bibliographystyle{apacite}
\bibliography{references}

\begin{thebibliography}{}

\bibitem [\protect \citeauthoryear {%
Alperin%
, Portenoy%
, Demes%
, Larivière%
\BCBL {}\ \BBA {} Haustein%
}{%
Alperin%
\ \protect \BOthers {.}}{%
{\protect \APACyear {2024}}%
}]{%
alperin_analysis_2024}
\APACinsertmetastar {%
alperin_analysis_2024}%
\begin{APACrefauthors}%
Alperin, J\BPBI P.%
, Portenoy, J.%
, Demes, K.%
, Larivière, V.%
\BCBL {}\ \BBA {} Haustein, S.%
\end{APACrefauthors}%
\unskip\
\newblock
\APACrefYearMonthDay{2024}{}{}.
\newblock
\APACrefbtitle {An analysis of the suitability of {OpenAlex} for bibliometric analyses.} {An analysis of the suitability of {OpenAlex} for bibliometric analyses.}
\newblock
\APACaddressPublisher{}{arXiv}.
\newblock
\begin{APACrefURL} [{2024-05-06}]\url{http://arxiv.org/abs/2404.17663} \end{APACrefURL}
\newblock
\begin{APACrefDOI} \doi{10.48550/arXiv.2404.17663} \end{APACrefDOI}
\PrintBackRefs{\CurrentBib}

\bibitem [\protect \citeauthoryear {%
Bashir%
, Surian%
\BCBL {}\ \BBA {} Dunn%
}{%
Bashir%
\ \protect \BOthers {.}}{%
{\protect \APACyear {2018}}%
}]{%
bashir_sr_update_2018}
\APACinsertmetastar {%
bashir_sr_update_2018}%
\begin{APACrefauthors}%
Bashir, R.%
, Surian, D.%
\BCBL {}\ \BBA {} Dunn, A\BPBI G.%
\end{APACrefauthors}%
\unskip\
\newblock
\APACrefYearMonthDay{2018}{}{}.
\newblock
{\BBOQ}\APACrefatitle {Time-to-update of systematic reviews relative to the availability of new evidence} {Time-to-update of systematic reviews relative to the availability of new evidence}.{\BBCQ}
\newblock
\APACjournalVolNumPages{Systematic Reviews}{7}{1}{195}.
\newblock
\begin{APACrefDOI} \doi{10.1186/s13643-018-0856-9} \end{APACrefDOI}
\PrintBackRefs{\CurrentBib}

\bibitem [\protect \citeauthoryear {%
Charalampous%
\ \BBA {} Knoth%
}{%
Charalampous%
\ \BBA {} Knoth%
}{%
{\protect \APACyear {2017}}%
}]{%
charalampous_classifying_2017}
\APACinsertmetastar {%
charalampous_classifying_2017}%
\begin{APACrefauthors}%
Charalampous, A.%
\BCBT {}\ \BBA {} Knoth, P.%
\end{APACrefauthors}%
\unskip\
\newblock
\APACrefYearMonthDay{2017}{}{}.
\newblock
\APACrefbtitle {Classifying document types to enhance search and recommendations in digital libraries.} {Classifying document types to enhance search and recommendations in digital libraries.}
\newblock
\APACaddressPublisher{}{arXiv}.
\newblock
\begin{APACrefURL} [{2025-01-09}]\url{http://arxiv.org/abs/1707.04134} \end{APACrefURL}
\newblock
\begin{APACrefDOI} \doi{10.48550/arXiv.1707.04134} \end{APACrefDOI}
\PrintBackRefs{\CurrentBib}

\bibitem [\protect \citeauthoryear {%
Culbert%
\ \protect \BOthers {.}}{%
Culbert%
\ \protect \BOthers {.}}{%
{\protect \APACyear {2025}}%
}]{%
culbert_reference_2025}
\APACinsertmetastar {%
culbert_reference_2025}%
\begin{APACrefauthors}%
Culbert, J\BPBI H.%
, Hobert, A.%
, Jahn, N.%
, Haupka, N.%
, Schmidt, M.%
, Donner, P.%
\BCBL {}\ \BBA {} Mayr, P.%
\end{APACrefauthors}%
\unskip\
\newblock
\APACrefYearMonthDay{2025}{}{}.
\newblock
{\BBOQ}\APACrefatitle {Reference coverage analysis of {OpenAlex} compared to {Web} of {Science} and {Scopus}} {Reference coverage analysis of {OpenAlex} compared to {Web} of {Science} and {Scopus}}.{\BBCQ}
\newblock
\APACjournalVolNumPages{Scientometrics}{130}{4}{2475--2492}.
\newblock
\begin{APACrefURL} \url{https://doi.org/10.1007/s11192-025-05293-3} \end{APACrefURL}
\newblock
\begin{APACrefDOI} \doi{10.1007/s11192-025-05293-3} \end{APACrefDOI}
\PrintBackRefs{\CurrentBib}

\bibitem [\protect \citeauthoryear {%
Delgado-Quirós%
\ \protect \BOthers {.}}{%
Delgado-Quirós%
\ \protect \BOthers {.}}{%
{\protect \APACyear {2024}}%
}]{%
delgado-quiros_why_2024}
\APACinsertmetastar {%
delgado-quiros_why_2024}%
\begin{APACrefauthors}%
Delgado-Quirós, L.%
, Aguillo, I\BPBI F.%
, Martín-Martín, A.%
, López-Cózar, E\BPBI D.%
, Orduña-Malea, E.%
\BCBL {}\ \BBA {} Ortega, J\BPBI L.%
\end{APACrefauthors}%
\unskip\
\newblock
\APACrefYearMonthDay{2024}{}{}.
\newblock
{\BBOQ}\APACrefatitle {Why are these publications missing? {Uncovering} the reasons behind the exclusion of documents in free-access scholarly databases} {Why are these publications missing? {Uncovering} the reasons behind the exclusion of documents in free-access scholarly databases}.{\BBCQ}
\newblock
\APACjournalVolNumPages{Journal of the Association for Information Science and Technology}{75}{1}{43--58}.
\newblock
\begin{APACrefURL} [{2025-05-30}]\url{https://onlinelibrary.wiley.com/doi/abs/10.1002/asi.24839} \end{APACrefURL}
\newblock
\begin{APACrefDOI} \doi{10.1002/asi.24839} \end{APACrefDOI}
\PrintBackRefs{\CurrentBib}

\bibitem [\protect \citeauthoryear {%
Delgado-Quirós%
\ \BBA {} Ortega%
}{%
Delgado-Quirós%
\ \BBA {} Ortega%
}{%
{\protect \APACyear {2024}}%
}]{%
delgado-quiros_completeness_2024}
\APACinsertmetastar {%
delgado-quiros_completeness_2024}%
\begin{APACrefauthors}%
Delgado-Quirós, L.%
\BCBT {}\ \BBA {} Ortega, J\BPBI L.%
\end{APACrefauthors}%
\unskip\
\newblock
\APACrefYearMonthDay{2024}{}{}.
\newblock
{\BBOQ}\APACrefatitle {Completeness degree of publication metadata in eight free-access scholarly databases} {Completeness degree of publication metadata in eight free-access scholarly databases}.{\BBCQ}
\newblock
\APACjournalVolNumPages{Quantitative Science Studies}{}{}{1--19}.
\newblock
\begin{APACrefURL} \url{https://direct.mit.edu/qss/article/doi/10.1162/qss_a_00286/119466/Completeness-degree-of-publication-metadata-in} \end{APACrefURL}
\newblock
\begin{APACrefDOI} \doi{10.1162/qss_a_00286} \end{APACrefDOI}
\PrintBackRefs{\CurrentBib}

\bibitem [\protect \citeauthoryear {%
Donner%
}{%
Donner%
}{%
{\protect \APACyear {2017}}%
}]{%
donner_document_2017}
\APACinsertmetastar {%
donner_document_2017}%
\begin{APACrefauthors}%
Donner, P.%
\end{APACrefauthors}%
\unskip\
\newblock
\APACrefYearMonthDay{2017}{}{}.
\newblock
{\BBOQ}\APACrefatitle {Document type assignment accuracy in the journal citation index data of {Web} of {Science}} {Document type assignment accuracy in the journal citation index data of {Web} of {Science}}.{\BBCQ}
\newblock
\APACjournalVolNumPages{Scientometrics}{113}{1}{219--236}.
\newblock
\begin{APACrefURL} \url{https://doi.org/10.1007/s11192-017-2483-y} \end{APACrefURL}
\newblock
\begin{APACrefDOI} \doi{10.1007/s11192-017-2483-y} \end{APACrefDOI}
\PrintBackRefs{\CurrentBib}

\bibitem [\protect \citeauthoryear {%
Harzing%
}{%
Harzing%
}{%
{\protect \APACyear {2013}}%
}]{%
harzing_document_2013}
\APACinsertmetastar {%
harzing_document_2013}%
\begin{APACrefauthors}%
Harzing, A\BHBI W.%
\end{APACrefauthors}%
\unskip\
\newblock
\APACrefYearMonthDay{2013}{}{}.
\newblock
{\BBOQ}\APACrefatitle {Document categories in the {ISI} {Web} of {Knowledge}: {Misunderstanding} the {Social} {Sciences}?} {Document categories in the {ISI} {Web} of {Knowledge}: {Misunderstanding} the {Social} {Sciences}?}{\BBCQ}
\newblock
\APACjournalVolNumPages{Scientometrics}{94}{1}{23--34}.
\newblock
\begin{APACrefURL} [{2020-07-01}]\url{http://link.springer.com/10.1007/s11192-012-0738-1} \end{APACrefURL}
\newblock
\begin{APACrefDOI} \doi{10.1007/s11192-012-0738-1} \end{APACrefDOI}
\PrintBackRefs{\CurrentBib}

\bibitem [\protect \citeauthoryear {%
Haupka%
, Dörner%
\BCBL {}\ \BBA {} Jahn%
}{%
Haupka%
\ \protect \BOthers {.}}{%
{\protect \APACyear {2024}}%
}]{%
haupka_recent_2024}
\APACinsertmetastar {%
haupka_recent_2024}%
\begin{APACrefauthors}%
Haupka, N.%
, Dörner, S.%
\BCBL {}\ \BBA {} Jahn, N.%
\end{APACrefauthors}%
\unskip\
\newblock
\APACrefYearMonthDay{2024}{}{}.
\newblock
\APACrefbtitle {Recent {Changes} in {Document} type classification in {OpenAlex} compared to {Web} of {Science} and {Scopus}.} {Recent {Changes} in {Document} type classification in {OpenAlex} compared to {Web} of {Science} and {Scopus}.}
\newblock
\begin{APACrefURL} [{2025-01-22}]\url{https://subugoe.github.io/scholcomm_analytics/posts/openalex_document_types/} \end{APACrefURL}
\PrintBackRefs{\CurrentBib}

\bibitem [\protect \citeauthoryear {%
Hauschke%
\ \BBA {} Nazarovets%
}{%
Hauschke%
\ \BBA {} Nazarovets%
}{%
{\protect \APACyear {2025}}%
}]{%
hauschke_non-retracted_2025}
\APACinsertmetastar {%
hauschke_non-retracted_2025}%
\begin{APACrefauthors}%
Hauschke, C.%
\BCBT {}\ \BBA {} Nazarovets, S.%
\end{APACrefauthors}%
\unskip\
\newblock
\APACrefYearMonthDay{2025}{}{}.
\newblock
{\BBOQ}\APACrefatitle {({Non}-)retracted academic papers in {OpenAlex}} {({Non}-)retracted academic papers in {OpenAlex}}.{\BBCQ}
\newblock
\APACjournalVolNumPages{Journal of Information Science}{}{}{01655515251322478}.
\newblock
\begin{APACrefURL} \url{https://doi.org/10.1177/01655515251322478} \end{APACrefURL}
\newblock
\begin{APACrefDOI} \doi{10.1177/01655515251322478} \end{APACrefDOI}
\PrintBackRefs{\CurrentBib}

\bibitem [\protect \citeauthoryear {%
Hendricks%
, Tkaczyk%
, Lin%
\BCBL {}\ \BBA {} Feeney%
}{%
Hendricks%
\ \protect \BOthers {.}}{%
{\protect \APACyear {2020}}%
}]{%
hendricks_crossref_2020}
\APACinsertmetastar {%
hendricks_crossref_2020}%
\begin{APACrefauthors}%
Hendricks, G.%
, Tkaczyk, D.%
, Lin, J.%
\BCBL {}\ \BBA {} Feeney, P.%
\end{APACrefauthors}%
\unskip\
\newblock
\APACrefYearMonthDay{2020}{}{}.
\newblock
{\BBOQ}\APACrefatitle {Crossref: {The} sustainable source of community-owned scholarly metadata} {Crossref: {The} sustainable source of community-owned scholarly metadata}.{\BBCQ}
\newblock
\APACjournalVolNumPages{Quantitative Science Studies}{1}{1}{414--427}.
\newblock
\begin{APACrefURL} [{2023-04-19}]\url{https://doi.org/10.1162/qss_a_00022} \end{APACrefURL}
\newblock
\begin{APACrefDOI} \doi{10.1162/qss_a_00022} \end{APACrefDOI}
\PrintBackRefs{\CurrentBib}

\bibitem [\protect \citeauthoryear {%
Jiao%
, Li%
\BCBL {}\ \BBA {} Fang%
}{%
Jiao%
\ \protect \BOthers {.}}{%
{\protect \APACyear {2023}}%
}]{%
jiao_how_2023}
\APACinsertmetastar {%
jiao_how_2023}%
\begin{APACrefauthors}%
Jiao, C.%
, Li, K.%
\BCBL {}\ \BBA {} Fang, Z.%
\end{APACrefauthors}%
\unskip\
\newblock
\APACrefYearMonthDay{2023}{}{}.
\newblock
{\BBOQ}\APACrefatitle {How are exclusively data journals indexed in major scholarly databases? {An} examination of four databases} {How are exclusively data journals indexed in major scholarly databases? {An} examination of four databases}.{\BBCQ}
\newblock
\APACjournalVolNumPages{Scientific Data}{10}{1}{737}.
\newblock
\begin{APACrefURL} [{2025-03-21}]\url{https://www.nature.com/articles/s41597-023-02625-x} \end{APACrefURL}
\newblock
\begin{APACrefDOI} \doi{10.1038/s41597-023-02625-x} \end{APACrefDOI}
\PrintBackRefs{\CurrentBib}

\bibitem [\protect \citeauthoryear {%
Kinney%
\ \protect \BOthers {.}}{%
Kinney%
\ \protect \BOthers {.}}{%
{\protect \APACyear {2023}}%
}]{%
kinney_semantic_2023}
\APACinsertmetastar {%
kinney_semantic_2023}%
\begin{APACrefauthors}%
Kinney, R.%
, Anastasiades, C.%
, Authur, R.%
, Beltagy, I.%
, Bragg, J.%
, Buraczynski, A.%
\BDBL {}Weld, D\BPBI S.%
\end{APACrefauthors}%
\unskip\
\newblock
\APACrefYearMonthDay{2023}{}{}.
\newblock
\APACrefbtitle {The {Semantic} {Scholar} {Open} {Data} {Platform}.} {The {Semantic} {Scholar} {Open} {Data} {Platform}.}
\newblock
\begin{APACrefURL} [{2024-06-12}]\url{http://arxiv.org/abs/2301.10140} \end{APACrefURL}
\newblock
\begin{APACrefDOI} \doi{10.48550/arXiv.2301.10140} \end{APACrefDOI}
\PrintBackRefs{\CurrentBib}

\bibitem [\protect \citeauthoryear {%
Maddi%
, Maisonobe%
\BCBL {}\ \BBA {} Boukacem-Zeghmouri%
}{%
Maddi%
\ \protect \BOthers {.}}{%
{\protect \APACyear {2025}}%
}]{%
maddi_geographical_2025}
\APACinsertmetastar {%
maddi_geographical_2025}%
\begin{APACrefauthors}%
Maddi, A.%
, Maisonobe, M.%
\BCBL {}\ \BBA {} Boukacem-Zeghmouri, C.%
\end{APACrefauthors}%
\unskip\
\newblock
\APACrefYearMonthDay{2025}{}{}.
\newblock
{\BBOQ}\APACrefatitle {Geographical and disciplinary coverage of open access journals: {OpenAlex}, {Scopus}, and {WoS}} {Geographical and disciplinary coverage of open access journals: {OpenAlex}, {Scopus}, and {WoS}}.{\BBCQ}
\newblock
\APACjournalVolNumPages{PLOS ONE}{20}{4}{e0320347}.
\newblock
\begin{APACrefURL} [{2025-06-02}]\url{https://journals.plos.org/plosone/article?id=10.1371/journal.pone.0320347} \end{APACrefURL}
\newblock
\begin{APACrefDOI} \doi{10.1371/journal.pone.0320347} \end{APACrefDOI}
\PrintBackRefs{\CurrentBib}

\bibitem [\protect \citeauthoryear {%
Maisano%
, Mastrogiacomo%
, Ferrara%
\BCBL {}\ \BBA {} Franceschini%
}{%
Maisano%
\ \protect \BOthers {.}}{%
{\protect \APACyear {2025}}%
}]{%
maisano_large-scale_2025}
\APACinsertmetastar {%
maisano_large-scale_2025}%
\begin{APACrefauthors}%
Maisano, D\BPBI A.%
, Mastrogiacomo, L.%
, Ferrara, L.%
\BCBL {}\ \BBA {} Franceschini, F.%
\end{APACrefauthors}%
\unskip\
\newblock
\APACrefYearMonthDay{2025}{}{}.
\newblock
{\BBOQ}\APACrefatitle {A large-scale semi-automated approach for assessing document-type classification errors in bibliometric databases} {A large-scale semi-automated approach for assessing document-type classification errors in bibliometric databases}.{\BBCQ}
\newblock
\APACjournalVolNumPages{Scientometrics}{}{}{}.
\newblock
\begin{APACrefURL} [{2025-04-08}]\url{https://doi.org/10.1007/s11192-025-05244-y} \end{APACrefURL}
\newblock
\begin{APACrefDOI} \doi{10.1007/s11192-025-05244-y} \end{APACrefDOI}
\PrintBackRefs{\CurrentBib}

\bibitem [\protect \citeauthoryear {%
Moed%
\ \BBA {} Van~Leeuwen%
}{%
Moed%
\ \BBA {} Van~Leeuwen%
}{%
{\protect \APACyear {1995}}%
}]{%
moed_improving_1995}
\APACinsertmetastar {%
moed_improving_1995}%
\begin{APACrefauthors}%
Moed, H\BPBI F.%
\BCBT {}\ \BBA {} Van~Leeuwen, T\BPBI N.%
\end{APACrefauthors}%
\unskip\
\newblock
\APACrefYearMonthDay{1995}{}{}.
\newblock
{\BBOQ}\APACrefatitle {Improving the {Accuracy} of the {Institute} for {Scientific} {Information}'s {Journal} {Impact} {Factors}} {Improving the {Accuracy} of the {Institute} for {Scientific} {Information}'s {Journal} {Impact} {Factors}}.{\BBCQ}
\newblock
\APACjournalVolNumPages{Journal of the American Society for Information Science}{46}{6}{461--67}.
\PrintBackRefs{\CurrentBib}

\bibitem [\protect \citeauthoryear {%
Mokhnacheva%
}{%
Mokhnacheva%
}{%
{\protect \APACyear {2023}}%
}]{%
mokhnacheva_document_2023}
\APACinsertmetastar {%
mokhnacheva_document_2023}%
\begin{APACrefauthors}%
Mokhnacheva, Y\BPBI V.%
\end{APACrefauthors}%
\unskip\
\newblock
\APACrefYearMonthDay{2023}{}{}.
\newblock
{\BBOQ}\APACrefatitle {Document {Types} {Indexed} in {WoS} and {Scopus}: {Similarities}, {Differences}, and {Their} {Significance} in the {Analysis} of {Publication} {Activity}} {Document {Types} {Indexed} in {WoS} and {Scopus}: {Similarities}, {Differences}, and {Their} {Significance} in the {Analysis} of {Publication} {Activity}}.{\BBCQ}
\newblock
\APACjournalVolNumPages{Scientific and Technical Information Processing}{50}{1}{40--46}.
\newblock
\begin{APACrefURL} \url{https://doi.org/10.3103/S0147688223010033} \end{APACrefURL}
\newblock
\begin{APACrefDOI} \doi{10.3103/S0147688223010033} \end{APACrefDOI}
\PrintBackRefs{\CurrentBib}

\bibitem [\protect \citeauthoryear {%
Mongeon%
\ \protect \BOthers {.}}{%
Mongeon%
\ \protect \BOthers {.}}{%
{\protect \APACyear {2025}}%
}]{%
mongeon_investigating_2025}
\APACinsertmetastar {%
mongeon_investigating_2025}%
\begin{APACrefauthors}%
Mongeon, P.%
, Hare, M.%
, Krause, G.%
, Marjoram, R.%
, Riddle, P.%
, Toupin, R.%
\BCBL {}\ \BBA {} Wilson, S.%
\end{APACrefauthors}%
\unskip\
\newblock
\APACrefYearMonthDay{2025}{}{}.
\newblock
{\BBOQ}\APACrefatitle {Investigating {Document} {Type} {Discrepancies} between {OpenAlex} and the {Web} of {Science}} {Investigating {Document} {Type} {Discrepancies} between {OpenAlex} and the {Web} of {Science}}.{\BBCQ}
\newblock
\APACjournalVolNumPages{Proceedings of the Annual Conference of CAIS / Actes du congrès annuel de l'ACSI}{}{}{}.
\newblock
\begin{APACrefURL} [{2025-05-30}]\url{https://journals.library.ualberta.ca/ojs.cais-acsi.ca/index.php/cais-asci/article/view/1943} \end{APACrefURL}
\newblock
\begin{APACrefDOI} \doi{10.29173/cais1943} \end{APACrefDOI}
\PrintBackRefs{\CurrentBib}

\bibitem [\protect \citeauthoryear {%
Mork%
, Yepes%
\BCBL {}\ \BBA {} Aronson%
}{%
Mork%
\ \protect \BOthers {.}}{%
{\protect \APACyear {2013}}%
}]{%
mork_nlm_2013}
\APACinsertmetastar {%
mork_nlm_2013}%
\begin{APACrefauthors}%
Mork, J\BPBI G.%
, Yepes, A\BPBI J\BPBI J.%
\BCBL {}\ \BBA {} Aronson, A\BPBI R.%
\end{APACrefauthors}%
\unskip\
\newblock
\APACrefYearMonthDay{2013}{}{}.
\newblock
\APACrefbtitle {The {NLM} {Medical} {Text} {Indexer} {System} for {Indexing} {Biomedical} {Literature}.} {The {NLM} {Medical} {Text} {Indexer} {System} for {Indexing} {Biomedical} {Literature}.}
\newblock
\APACaddressPublisher{}{Lister Hill National Center for Biomedical Communications}.
\newblock
\begin{APACrefURL} \url{https://lhncbc.nlm.nih.gov/ii/information/Papers/MTI_System_Description_Expanded_2013_Accessible.pdf} \end{APACrefURL}
\PrintBackRefs{\CurrentBib}

\bibitem [\protect \citeauthoryear {%
Ortega%
}{%
Ortega%
}{%
{\protect \APACyear {2022}}%
}]{%
whisperer_when_2022}
\APACinsertmetastar {%
whisperer_when_2022}%
\begin{APACrefauthors}%
Ortega, J\BPBI L.%
\end{APACrefauthors}%
\unskip\
\newblock
\APACrefYearMonthDay{2022}{}{}.
\newblock
\APACrefbtitle {When is a paper published?} {When is a paper published?}
\newblock
\begin{APACrefURL} [{2024-03-28}]\url{https://researchwhisperer.org/2022/02/08/when-is-a-paper-published/} \end{APACrefURL}
\PrintBackRefs{\CurrentBib}

\bibitem [\protect \citeauthoryear {%
Priem%
, Piwowar%
\BCBL {}\ \BBA {} Orr%
}{%
Priem%
\ \protect \BOthers {.}}{%
{\protect \APACyear {2022}}%
}]{%
priem_openalex_2022}
\APACinsertmetastar {%
priem_openalex_2022}%
\begin{APACrefauthors}%
Priem, J.%
, Piwowar, H.%
\BCBL {}\ \BBA {} Orr, R.%
\end{APACrefauthors}%
\unskip\
\newblock
\APACrefYearMonthDay{2022}{}{}.
\newblock
\APACrefbtitle {{OpenAlex}: {A} fully-open index of scholarly works, authors, venues, institutions, and concepts.} {{OpenAlex}: {A} fully-open index of scholarly works, authors, venues, institutions, and concepts.}
\newblock
\APACaddressPublisher{}{arXiv}.
\newblock
\begin{APACrefURL} [{2024-04-10}]\url{http://arxiv.org/abs/2205.01833} \end{APACrefURL}
\newblock
\begin{APACrefDOI} \doi{10.48550/arXiv.2205.01833} \end{APACrefDOI}
\PrintBackRefs{\CurrentBib}

\bibitem [\protect \citeauthoryear {%
Scheidsteger%
\ \BBA {} Haunschild%
}{%
Scheidsteger%
\ \BBA {} Haunschild%
}{%
{\protect \APACyear {2023}}%
}]{%
scheidsteger_comparison_2023}
\APACinsertmetastar {%
scheidsteger_comparison_2023}%
\begin{APACrefauthors}%
Scheidsteger, T.%
\BCBT {}\ \BBA {} Haunschild, R.%
\end{APACrefauthors}%
\unskip\
\newblock
\APACrefYearMonthDay{2023}{}{}.
\newblock
{\BBOQ}\APACrefatitle {Comparison of metadata with relevance for bibliometrics between {Microsoft} {Academic} {Graph} and {OpenAlex} until 2020} {Comparison of metadata with relevance for bibliometrics between {Microsoft} {Academic} {Graph} and {OpenAlex} until 2020}.{\BBCQ}
\newblock
\APACjournalVolNumPages{El Profesional de la información}{}{}{}.
\newblock
\begin{APACrefURL} \url{http://arxiv.org/abs/2206.14168} \end{APACrefURL}
\newblock
\begin{APACrefDOI} \doi{10.3145/epi.2023.mar.09} \end{APACrefDOI}
\PrintBackRefs{\CurrentBib}

\bibitem [\protect \citeauthoryear {%
Schmidt%
\ \protect \BOthers {.}}{%
Schmidt%
\ \protect \BOthers {.}}{%
{\protect \APACyear {2024}}%
}]{%
schmidt_data_2024}
\APACinsertmetastar {%
schmidt_data_2024}%
\begin{APACrefauthors}%
Schmidt, M.%
, Rimmert, C.%
, Stephen, D.%
, Lenke, C.%
, Donner, P.%
, Gärtner, S.%
\BDBL {}Stahlschmidt, S.%
\end{APACrefauthors}%
\unskip\
\newblock
\APACrefYearMonthDay{2024}{}{}.
\newblock
\APACrefbtitle {The {Data} {Infrastructure} of the {German} {Kompetenznetzwerk} {Bibliometrie}: {An} {Enabling} {Intermediary} between {Raw} {Data} and {Analysis}.} {The {Data} {Infrastructure} of the {German} {Kompetenznetzwerk} {Bibliometrie}: {An} {Enabling} {Intermediary} between {Raw} {Data} and {Analysis}.}
\newblock
\APACaddressPublisher{}{Zenodo}.
\newblock
\begin{APACrefURL} [{2025-01-22}]\url{https://zenodo.org/records/13935407} \end{APACrefURL}
\newblock
\begin{APACrefDOI} \doi{10.5281/zenodo.13935407} \end{APACrefDOI}
\PrintBackRefs{\CurrentBib}

\bibitem [\protect \citeauthoryear {%
Sigogneau%
}{%
Sigogneau%
}{%
{\protect \APACyear {2000}}%
}]{%
sigogneau_2020}
\APACinsertmetastar {%
sigogneau_2020}%
\begin{APACrefauthors}%
Sigogneau, A.%
\end{APACrefauthors}%
\unskip\
\newblock
\APACrefYearMonthDay{2000}{}{}.
\newblock
{\BBOQ}\APACrefatitle {An analysis of document types published in journals related to physics: Proceeding papers recorded in the Science Citation Index database} {An analysis of document types published in journals related to physics: Proceeding papers recorded in the science citation index database}.{\BBCQ}
\newblock
\APACjournalVolNumPages{Scientometrics}{47}{3}{589--604}.
\newblock
\begin{APACrefURL} \url{https://link.springer.com/article/10.1023/A:1005628218890} \end{APACrefURL}
\newblock
\begin{APACrefDOI} \doi{https://doi.org/10.1023/A:1005628218890} \end{APACrefDOI}
\PrintBackRefs{\CurrentBib}

\bibitem [\protect \citeauthoryear {%
Torre%
}{%
Torre%
}{%
{\protect \APACyear {2012}}%
}]{%
torre_versioning_2012}
\APACinsertmetastar {%
torre_versioning_2012}%
\begin{APACrefauthors}%
Torre, S.%
\end{APACrefauthors}%
\unskip\
\newblock
\APACrefYearMonthDay{2012}{}{}.
\newblock
{\BBOQ}\APACrefatitle {Versioning in {PubMed}} {Versioning in {PubMed}}.{\BBCQ}
\newblock
\APACjournalVolNumPages{NLM Tech Bull}{}{384}{e6}.
\newblock
\begin{APACrefURL} \url{https://www.nlm.nih.gov/pubs/techbull/jf12/jf12_pm_versioning.html} \end{APACrefURL}
\PrintBackRefs{\CurrentBib}

\bibitem [\protect \citeauthoryear {%
van Buskirk%
}{%
van Buskirk%
}{%
{\protect \APACyear {1984}}%
}]{%
van_buskirk_review_1984}
\APACinsertmetastar {%
van_buskirk_review_1984}%
\begin{APACrefauthors}%
van Buskirk, N\BPBI E.%
\end{APACrefauthors}%
\unskip\
\newblock
\APACrefYearMonthDay{1984}{}{}.
\newblock
{\BBOQ}\APACrefatitle {The {Review} {Article} in {MEDLINE}: {Ambiguity} of {Definition} and {Implications} for {Online} {Searchers}} {The {Review} {Article} in {MEDLINE}: {Ambiguity} of {Definition} and {Implications} for {Online} {Searchers}}.{\BBCQ}
\newblock
\APACjournalVolNumPages{Bull. Med. Libr. Assoc.}{72}{4}{349--352}.
\PrintBackRefs{\CurrentBib}

\bibitem [\protect \citeauthoryear {%
Van~Eck%
\ \BBA {} Waltman%
}{%
Van~Eck%
\ \BBA {} Waltman%
}{%
{\protect \APACyear {2024}}%
}]{%
van_eck_2024_13879947}
\APACinsertmetastar {%
van_eck_2024_13879947}%
\begin{APACrefauthors}%
Van~Eck, N\BPBI J.%
\BCBT {}\ \BBA {} Waltman, L.%
\end{APACrefauthors}%
\unskip\
\newblock
\APACrefYearMonthDay{2024}{}{}.
\newblock
\APACrefbtitle {A methodology for identifying core sources and core publications in OpenAlex.} {A methodology for identifying core sources and core publications in openalex.}
\newblock
\APACaddressPublisher{}{Zenodo}.
\newblock
\begin{APACrefURL} \url{https://doi.org/10.5281/zenodo.13879947} \end{APACrefURL}
\newblock
\begin{APACrefDOI} \doi{10.5281/zenodo.13879947} \end{APACrefDOI}
\PrintBackRefs{\CurrentBib}

\bibitem [\protect \citeauthoryear {%
van Raan%
}{%
van Raan%
}{%
{\protect \APACyear {2004}}%
}]{%
moed_measuring_2004}
\APACinsertmetastar {%
moed_measuring_2004}%
\begin{APACrefauthors}%
van Raan, A\BPBI F\BPBI J.%
\end{APACrefauthors}%
\unskip\
\newblock
\APACrefYearMonthDay{2004}{}{}.
\newblock
{\BBOQ}\APACrefatitle {Measuring {Science}} {Measuring {Science}}.{\BBCQ}
\newblock
\BIn{} H\BPBI F.~Moed, W.~Glänzel\BCBL {}\ \BBA {} U.~Schmoch\ (\BEDS), \APACrefbtitle {Handbook of {Quantitative} {Science} and {Technology} {Research}} {Handbook of {Quantitative} {Science} and {Technology} {Research}}\ (\BPGS\ 19--50).
\newblock
\APACaddressPublisher{Dordrecht}{Springer Netherlands}.
\newblock
\begin{APACrefURL} \url{http://link.springer.com/10.1007/1-4020-2755-9} \end{APACrefURL}
\newblock
\begin{APACrefDOI} \doi{10.1007/1-4020-2755-9} \end{APACrefDOI}
\PrintBackRefs{\CurrentBib}

\bibitem [\protect \citeauthoryear {%
Visser%
, van Eck%
\BCBL {}\ \BBA {} Waltman%
}{%
Visser%
\ \protect \BOthers {.}}{%
{\protect \APACyear {2021}}%
}]{%
visser_large-scale_2021}
\APACinsertmetastar {%
visser_large-scale_2021}%
\begin{APACrefauthors}%
Visser, M.%
, van Eck, N\BPBI J.%
\BCBL {}\ \BBA {} Waltman, L.%
\end{APACrefauthors}%
\unskip\
\newblock
\APACrefYearMonthDay{2021}{}{}.
\newblock
{\BBOQ}\APACrefatitle {Large-scale comparison of bibliographic data sources: {Scopus}, {Web} of {Science}, {Dimensions}, {Crossref}, and {Microsoft} {Academic}} {Large-scale comparison of bibliographic data sources: {Scopus}, {Web} of {Science}, {Dimensions}, {Crossref}, and {Microsoft} {Academic}}.{\BBCQ}
\newblock
\APACjournalVolNumPages{Quantitative Science Studies}{2}{1}{20--41}.
\newblock
\begin{APACrefURL} \url{https://doi.org/10.1162/qss\_a\_00112} \end{APACrefURL}
\newblock
\begin{APACrefDOI} \doi{10.1162/qss_a_00112} \end{APACrefDOI}
\PrintBackRefs{\CurrentBib}

\bibitem [\protect \citeauthoryear {%
Waltman%
\ \BBA {} Larivière%
}{%
Waltman%
\ \BBA {} Larivière%
}{%
{\protect \APACyear {2020}}%
}]{%
waltman_special_2020}
\APACinsertmetastar {%
waltman_special_2020}%
\begin{APACrefauthors}%
Waltman, L.%
\BCBT {}\ \BBA {} Larivière, V.%
\end{APACrefauthors}%
\unskip\
\newblock
\APACrefYearMonthDay{2020}{}{}.
\newblock
{\BBOQ}\APACrefatitle {Special issue on bibliographic data sources} {Special issue on bibliographic data sources}.{\BBCQ}
\newblock
\APACjournalVolNumPages{Quantitative Science Studies}{1}{1}{360--362}.
\newblock
\begin{APACrefURL} \url{https://doi.org/10.1162/qss\_e\_00026} \end{APACrefURL}
\newblock
\begin{APACrefDOI} \doi{10.1162/qss_e_00026} \end{APACrefDOI}
\PrintBackRefs{\CurrentBib}

\bibitem [\protect \citeauthoryear {%
Xie%
, Shen%
\BCBL {}\ \BBA {} Wang%
}{%
Xie%
\ \protect \BOthers {.}}{%
{\protect \APACyear {2021}}%
}]{%
xie_is_2021}
\APACinsertmetastar {%
xie_is_2021}%
\begin{APACrefauthors}%
Xie, B.%
, Shen, Z.%
\BCBL {}\ \BBA {} Wang, K.%
\end{APACrefauthors}%
\unskip\
\newblock
\APACrefYearMonthDay{2021}{}{}.
\newblock
\APACrefbtitle {Is preprint the future of science? {A} thirty year journey of online preprint services.} {Is preprint the future of science? {A} thirty year journey of online preprint services.}
\newblock
\APACaddressPublisher{}{arXiv}.
\newblock
\begin{APACrefURL} [{2024-04-24}]\url{http://arxiv.org/abs/2102.09066} \end{APACrefURL}
\PrintBackRefs{\CurrentBib}

\bibitem [\protect \citeauthoryear {%
Yeung%
}{%
Yeung%
}{%
{\protect \APACyear {2019}}%
}]{%
yeung_comparison_2019}
\APACinsertmetastar {%
yeung_comparison_2019}%
\begin{APACrefauthors}%
Yeung, A\BPBI W\BPBI K.%
\end{APACrefauthors}%
\unskip\
\newblock
\APACrefYearMonthDay{2019}{}{}.
\newblock
{\BBOQ}\APACrefatitle {Comparison between {Scopus}, {Web} of {Science}, {PubMed} and publishers for mislabelled review papers} {Comparison between {Scopus}, {Web} of {Science}, {PubMed} and publishers for mislabelled review papers}.{\BBCQ}
\newblock
\APACjournalVolNumPages{Current Science}{116}{11}{1909--1914}.
\newblock
\begin{APACrefURL} \url{https://www.jstor.org/stable/27138143} \end{APACrefURL}
\PrintBackRefs{\CurrentBib}

\end{thebibliography}

\newpage
\section{Appendix}
\subsection{Mappings}
The mappings used for matching document types to the categories \textit{research discourse} and \textit{editorial discourse} as mentioned in section \ref{sec:methodology} are listed here.

\begin{table}[h]
\centering
\makebox[\textwidth]{
\begin{tabularx}{1.2\textwidth}{l|X}
\toprule
{} & {\textbf{Types}} \\
\hline
Research Discourse &  article, journal-article \\
Editorial Discourse & erratum, editorial, letter, paratext \\
Not assigned & grant, book-chapter, dataset, book, other, reference-entry, dissertation, report, peer-review, standard, book-series \\
\bottomrule
\end{tabularx}}
\caption{OpenAlex Type Mapping}
\label{tab:oal_mapping}
\end{table}

\begin{table}[h]
\centering
\makebox[\textwidth]{
\begin{tabularx}{1.2\textwidth}{l|X}
\toprule
{} & {\textbf{Types}} \\
\hline
Research Discourse &  Review, Article, {Article in Press} \\
Editorial Discourse & Erratum, Editorial, Letter, Note \\
Not assigned & Conference Paper, Chapter, Short Survey, Book, Tombstone, Data Paper, Conference Review, Abstract Report, Business Article, pp, Report \\
\bottomrule
\end{tabularx}}
\caption{Scopus Type Mapping. The item type \enquote{pp} is an artefact from the Scopus XML raw files.}
\label{tab:scp_mapping}
\end{table}

\begin{table}[h]
\centering
\makebox[\textwidth]{
\begin{tabularx}{1.2\textwidth}{l|X}
\toprule
{} & {\textbf{Types}} \\
\hline
Research Discourse &  Review, MetaAnalysis, JournalArticle, Study, CaseReport, ClinicalTrial \\
Editorial Discourse & Editorial, News, LettersAndComments \\
Not assigned & Conference, Book, Dataset \\
\bottomrule
\end{tabularx}}
\caption{Semantic Scholar Type Mapping}
\label{tab:s2_mapping}
\end{table}

\begin{table}[h]
\centering
\makebox[\textwidth]{
\begin{tabularx}{1.2\textwidth}{l|X}
\toprule
{} & {\textbf{Types}} \\
\hline
Research Discourse &  Cochrane Systematic Review, Systematic Review, Meta-Analysis, Review, Case Reports, Randomized Controlled Trial, Clinical Trial, Clinical Trial, Phase II, Clinical Trial, Phase III, Clinical Trial, Phase I, Clinical Trial, Phase IV, Controlled Clinical Trial, Pragmatic Clinical Trial, Journal Article, Comparative Study, Multicenter Study, Observational Study, Evaluation Study, Historical Article, Validation Study, Clinical Study, Randomized Controlled Trial, Veterinary, Twin Study, Clinical Trial, Veterinary, Classical Article, Observational Study, Veterinary, Corrected and Republished Article, Adaptive Clinical Trial, Evaluation Studies, Validation Studies, "Research Support, Non-U.S. Govt", "Research Support, N.I.H., Extramural", "Research Support, U.S. Govt, Non-P.H.S.", "Research Support, U.S. Govt, P.H.S.", Preprint, "Research Support, N.I.H., Intramural", "Research Support, American Recovery and Reinvestment Act", Equivalence Trial \\
Editorial Discourse & Published Erratum, Retraction of Publication, Retracted Publication, Editorial, News, Letter, Comment, Introductory Journal Article, Newspaper Article \\
Not assigned & English Abstract, Video-Audio Media, Biography, Practice Guideline, Portrait, Congress, Clinical Trial Protocol, Interview, Personal Narrative, Consensus Development Conference, Overall, Patient Education Handout, Guideline, Dataset, Lecture, Address, Clinical Conference, Expression of Concern, Legal Case, Autobiography, Technical Report, Webcast, Bibliography, Festschrift, Consensus Development Conference, NIH, Interactive Tutorial, Scientific Integrity Review, Duplicate Publication, Directory,  Periodical Index, Dictionary, Legislation \\
\bottomrule
\end{tabularx}}
\caption{PubMed Type Mapping}
\label{tab:pm_mapping}
\end{table}

\begin{table}[h]
\centering
\makebox[\textwidth]{
\begin{tabularx}{1.2\textwidth}{l|X}
\toprule
{} & {\textbf{Types}} \\
\hline
Research Discourse &  Review, Article, Retracted Publication \\
Editorial Discourse & Correction, Retraction,  Item Withdrawal, Editorial Material, News Item, Letter, Note \\
Not assigned & Meeting Abstract, Book Review, Early Access, Biographical-Item, Book Chapter, Poetry, Reprint, Data Paper, Bibliography, Fiction, Creative Prose, Art Exhibit Review, Theater Review, Software Review, CC Meeting Heading, Record Review, Expression of Concern, Film Review, Music Performance Review, Music Score Review, TV Review, Radio Review, Excerpt, Database Review, Script, Hardware Review, Dance Performance Review, Book, Music Score, Chronology, Meeting Summary, Main Cite, Meeting-Abstract, Proceedings Paper \\
\bottomrule
\end{tabularx}}
\caption{WoS Type Mapping}
\label{tab:wos_mapping}
\end{table}

\begin{table}[h]
\centering
\makebox[\textwidth]{
\begin{tabularx}{\textwidth}{lX}
\toprule
{\textbf{Rank}} & {\textbf{PubMed types}}\\
\hline
10 (Retractions) & Published Erratum; Retraction of Publication; Retracted Publication\\
11 (Editorials) & Editorial*\\
12 (Letters) & Letter*\\
13 (News) & News*\\
20 (Reviews) & Cochrane Systematic Review; Systematic Review; Meta-Analysis; Review\\
30 (Case reports) & Case Reports\\
31 (Clinical trial) & Clinical Trial, Phase IV; Pragmatic Clinical Trial; Controlled Clinical Trial; Randomized Controlled Trial; Clinical Trial; Clinical Trial, Phase II; Clinical Trial, Phase III; Clinical Trial, Phase I\\
50 (Articles) & Journal Article*\\
90 (Funding information) & Research Support, American Recovery and Reinvestment Act; Research Support, U.S. Gov't, Non-P.H.S.; Research Support, N.I.H., Intramural; Research Support, U.S. Gov't, P.H.S.; Research Support, Non-U.S. Gov't; Research Support, N.I.H., Extramural \\
99 (other)& ...\\           
\bottomrule
\end{tabularx}}
*PubMed indexing rules define Editorials, Letters, News, and Articles as base types that can never co-occur
\caption{PubMed Type Hierarchy}
\label{tab:pm_hierarchy}
\end{table}

\end{document}